\documentstyle[twocolumn,twoside]{article}
\newcommand{\h}{\linebreak \hspace*{3ex}}
\newcommand{\hb}{\\ \hspace*{2ex}}

\begin{document}
\title{ACCRETION DISCS AROUND BLACK HOLES: DEVELOPEMENT OF THEORY}
\author{G.S.\,Bisnovatyi-Kogan$^1$ \\[2mm]
\begin{tabular}{l}
 $^1$ Space Research Institute, Russian Academy of Sciences\hb
 Profsoyuznaya 84/32, Moscow 117810 Russia, {\em gkogan@mx.iki.rssi.ru}
 \\[2mm]
\end{tabular}
% no tabular, if one affiliation only
}
\date{}
\maketitle

ABSTRACT.
Standard accretion disk theory is formulated which is based
on the local heat balance. The energy produced by a
turbulent viscous
heating is supposed to be emitted to the sides of the disc.
Sources of turbulence in the accretion disc are
connected with nonlinear hydrodynamic instability, convection,
and magnetic field.
In standard theory there are two branches of solution, optically thick, and
optically thin.
Advection in accretion disks is described by the differential
equations what makes the theory nonlocal.
Low-luminous optically thin accretion disc model with advection
at some suggestions may become advectively dominated, carrying almost all
the energy inside the black hole. The proper account of magnetic filed
in the process of accretion limits the energy advected into a black hole,
efficiency of accretion
should exceed $\sim 1/4$ of the standard accretion disk model
efficiency. \\[1mm]
{\bf Key words}: Stars: accretion discs; black holes. \\[2mm]

{\bf 1. Introduction}\\[1mm]

Accretion is a main source of energy in binary X-ray sources,
quasars and active galactic nuclei (AGN). The intensive
development of accretion theory began after discovery of X-ray sources
and quasars.
Accretion into stars is ended by a collision with an inner boundary,
which may be a stellar surface, or outer boundary of a magnetosphere
for strongly magnetized stars. All
gravitational energy of the falling matter is transformed into heat
and radiated outward.

In black holes matter is falling to the horizon,
from where no radiation arrives.
All luminosity is formed on the way to it. The efficiency of accretion
is not known from the beginning,
and depends on angular momentum of the
falling matter, and magnetic field embedded into it. It was first
shown by Schwartsman (1971) that during spherical accretion of
nonmagnetized gas the efficiency may be as small as $10^{-8}$ for
sufficiently low mass fluxes.
He had shown that presence of magnetic field in
the accreting matter increase the efficiency up to about $10\%$, and
account of heating of matter due to magnetic field annihilation
rises the efficiency up to about $30\%$
(Bisnovatyi-Kogan and Ruzmaikin, 1974)
In the case of a thin disc accretion, when matter has
large angular momentum, the efficiency is about $1/2$ of the efficiency
of accretion
into a star with a radius equal to the radius of the last stable orbit.
In the case of geometrically thick and
optically thin accretion discs the situation is approaching the case
of spherical symmetry, where presence of a magnetic field plays
a critical role.

Advection dominated accretion flow (ADAF) was suggested
by Narayan and Yu (1995), and used as a solution for
some astrophysical problems.
The suggestions underlying ADAF: ignorance of the magnetic field annihilation
in heating of a plasma flow, electron heating only due to
binary collisions with protons (ions) had been critically analyzed
in papers of
Bisnovatyi-Kogan and Lovelace (1997, 1999),
Bisnovatyi-Kogan (1999), Quataert (1997).
There are contradictions between ADAF model and observational
data in radioemission of elliptical galaxies (Di Matteo et al., 1999),
and X-ray emission of Seyfert galaxy NGC4258 (Cannizzo, 1998).
Account of processes connected with a small-scale magnetic field
in accretion flow, strongly restricts
solution. Namely, the efficiency of the accretion flow cannot become less
then about 1/4 of the standard accretion disc value.\\[2mm]

{\bf 2. Development of the standard model}\\[1mm]

Matter falling into a black hole forms a disc when its
angular momentum is sufficiently high. It happens when matter
comes from the
star companion in the binary, or from
a tidal disruption of the star which trajectory
approaches close to the black hole.
The first situation is observed in many galactic X-ray sources
(Cherepashchuk, 1996).
A tidal disruption happens in quasars and
active galactic nuclei (AGN) in the model of supermassive black
hole surrounded by a dense stellar cluster.

Equations of a standard accretion disk theory had been first formulated
by (Shakura, 1972); some corrections and generalization to general
relativity (GR) had been done by Novikov and Thorne (1973). Observational
aspects of accretion disks have been analyzed by Shakura and Sunyaev (1973).
Note, that all authors
of the accretion disc theory from USSR were students (N.I.Shakura)
or collaborators (I.D.Novikov and R.A.Sunyaev) of academician
Ya.B.Zeldovich, who was not among the authors, but whose influence
on them hardly could be overestimated.
The main idea of this theory is to describe a
geometrically thin non-self-gravitating disc of a mass $M_d$, much
smaller then the mass of a black hole $M$, by hydrodynamic
equations averaged over the disc thickness $2h$. \\[2mm]

{\it 2.1. Equations}\\[1mm]

The small thickness of the disc in comparison with its radius $h \ll r$
means small importance of the pressure gradient
$\nabla P$ in comparison with
gravity and inertia forces.
Radial equilibrium equation in a disc is a balance between the last two
forces with an angular velocity equals to
the keplerian one $\Omega=\Omega_K=\left(\frac{GM}{r^3}\right)^{1/2}$.
Note, that just before a last stable orbit around a black hole
this suggestion fails, but in the "standard"
accretion disc model this relation is supposed to hold
all over the disc, with an inner boundary at the last stable orbit.
The equilibrium equation in the vertical $z$-direction is determined by a
balance between the gravitational force and pressure gradient
$\frac{dP}{dz}=-\rho\frac{GMz}{r^3}$.
For a thin disc this differential equation is substituted by an
algebraic one, determining the half-thickness of the disc in the form

\begin{equation}
\label{ref1.3}
h \approx \frac{1}{\Omega_K}\left(2\frac{P}{\rho}\right)^{1/2}.
\end{equation}
The balance of angular momentum, related to the $\phi$ component of the
Euler equation has an integral in a stationary case, written as

\begin{equation}
\label{ref1.4}
\dot M(j-j_{in})=-2\pi r^2\,2ht_{r\phi},\quad t_{r\phi}=
\eta r\frac{d\Omega}{dr}.
\end{equation}
Here $j=v_{\phi}r=\Omega r^2$ is a specific angular momentum,
$t_{r\phi}$ is a component of the viscous stress tensor, $\dot M>0$ is a mass
flux per unit time into a black hole,
$j_{in}$ is equal to the specific angular
momentum of matter
falling into a black hole. In the standard theory the value of $j_{in}$
is determined from physical considerations.
For accretion
into a black hole it is suggested, that on the last stable orbit the
gradient of the angular velocity is zero, corresponding to zero
viscous momentum flux. In that case
$j_{in}=\Omega_K r_{in}^2,$
corresponding to the Keplerian angular momentum of the matter on the last
stable orbit.

The choice of the viscosity  coefficient is the most
speculative problem of the theory. In the laminar case at
low microscopic (atomic or plasma) viscosity the stationary
accretion disc must be very massive and thick, and before its formation
the matter is collected by disc leading to a small flux inside.
It contradicts to observations of X-ray binaries,
where a considerable matter flux along the accretion disc may be explained
only when viscosity coefficient is much larger.
It was suggested by Shakura (1972), that matter in the disc
is turbulent, what determines a turbulent viscous stress tensor,
parametrized by a pressure

\begin{equation}
\label{ref1.7}
t_{r\phi}=-\alpha\rho v_s^2 = -\alpha P,
\end{equation}
where $v_s$ is a sound speed in the matter.
This simple presentation comes out from a relation for a turbulent viscosity
coefficient $\eta_t\approx \rho v_t l$ with an average turbulent velocity
$v_t$ and mean free path of the turbulent element $l$. It follows from
the definition of $t_{r\phi}$ in (\ref{ref1.4}), when we take $l \approx h$
from (\ref{ref1.3})

\begin{equation}
\label{ref1.8}
t_{r\phi}=\rho v_t h r \frac{d\Omega}{dr} \approx \rho v_t v_s =-\alpha
\rho v_s^2,
\end{equation}
where a coefficient $\alpha<1$ is connecting the turbulent and sound speeds
$v_t=\alpha v_s$. Presentations of $t_{r\phi}$ in (\ref{ref1.7}) and
(\ref{ref1.8}) are equivalent, and only when the angular velocity
differs considerably from the Keplerian one the first relation to the
right in (\ref{ref1.8}) is more preferable. That does not appear
in the standard theory, but may happen when advective
terms are included.

Development of a turbulence in the accretion disc cannot be justified
simply, because a Keplerian disc is stable in linear approximation to
the development of perturbations. It was suggested by Ya.B.Zeldovich,
that in presence of very large Reynolds number
${\rm Re}=\frac{\rho v l}{\eta}$
the amplitude of perturbations at which nonlinear effects become important
is very low, so a turbulence may be developed due to
nonlinear instability even when the disc is stable in linear approximation.
Viscous stresses may arise from a magnetic field,
it was suggested by (Shakura, 1972), that magnetic stresses cannot
exceed the turbulent ones.
It was shown by Bisnovatyi-Kogan and Blinnikov (1976), that inner
regions of a highly luminous accretion discs where pressure is dominated
by radiation, are unstable to vertical convection. Development of this
convection produce a turbulence, needed for a high viscosity.

With alpha- prescription of viscosity the equation of angular
momentum conservation is written in the plane of a disc as

\begin{equation}
\label{ref1.9}
\dot M(j-j_{in})=
4\pi r^2 \alpha P_0 h.
\end{equation}
When angular velocity is far from Keplerian one the relation
(\ref{ref1.4}) is valid with a coefficient of a turbulent viscosity
$\eta=\alpha\rho_0 v_{s0} h$,
where values with the index "0" denote the plane of the disc.

In the standard theory a heat balance is local,  all
heat produced by viscosity in the ring between $r$ and $r+dr$ is
radiated through the sides of disc at the same $r$. The heat production
rate $Q_+$ related to the surface unit of the disc is written as

\begin{equation}
\label{ref1.11}
Q_+=h\,t_{r\phi}r\frac{d\Omega}{dr}=\frac{3}{8\pi}\dot M \frac{GM}{r^3}
\left(1-\frac{j_{in}}{j}\right).
\end{equation}
Heat losses by a disc depend on its optical depth. The standard disc
model (Shakura, 1972) considered a geometrically thin disc as an optically
thick in a vertical direction. That implies energy losses $Q_-$ from the disc
due to a radiative conductivity, after a substitution of
the differential equation
of a heat transfer by an algebraic relation

\begin{equation}
\label{ref1.12}
Q_- \approx \frac{4}{3} \frac{acT^4}{\kappa \Sigma}.
\end{equation}
Here $a$ is a constant of a radiation energy density,
$c$ is a speed of light,
$T$ is a temperature in the disc plane, $\kappa$ is a matter opacity,
and a surface density
$\Sigma=2\rho h$. Here and below
$\rho,\, T,\,P$ without the index "0" are related to the disc plane.
The heat balance equation is represented by a relation
$Q_+=Q_-$.
A continuity equation in the standard model
is used for finding of a radial velocity $v_r$

\begin{equation}
\label{ref1.14a}
v_r=\frac{\dot M}{4\pi rh\rho}=\frac{\dot M}{2\pi r\Sigma}.
\end{equation}
Completing these equations by an equation of state $P(\rho,T)$ and relation
for the opacity
$\kappa=\kappa(\rho, T)$ we get a full set of equations
for a standard disc model. For power low equations of state of an ideal gas
$P=P_g=\rho {\cal R} T$ (${\cal R}$ is a gas constant), or radiation pressure
$P=P_r=\frac{aT^4}{3}$, and opacity in the form of electron scattering
$\kappa_e$, or Karammers formulae $\kappa_k$,
the solution of a standard disc accretion theory is obtained
analytically.
Checking the suggestion of a large optical thickness confirms a
self-consistency of the model. Note that
solutions for different regions
of the disc with different equation of states and opacities are not matched
to each other. \\[2mm]

{\it 2.2. Optically thin solution}\\[1mm]

It was found by Shapiro et al. (1976) that there is another branch
of the solution for a disc structure with the same input parameters
$M,\,\dot M,\,\alpha$ which is also self-consistent but has a small
optical thickness.
That implies another equation of energy losses,
determined by a volume emission
$Q_- \approx q\, \rho\,h$.
The emissivity of the unit of a volume $q$ is
connected with a Planckian averaged opacity
$\kappa_p$ by a relation
$q \approx acT_0^4 \kappa_p$. In the optically thin limit the
pressure is determined by a gas $P=P_g$.
In the optically thin solution the thickness of the disc is larger then
in the optically thick one, and density is lower.

While heating
by viscosity is determined mainly by heavy ions, and cooling is determined
by electrons, the rate of the energy exchange between them is important for
a structure of the disc. The energy balance equations are written
separately for ions and electrons. For small accretion rates and lower
matter density the rate of energy exchange due to binary collisions is
so slow, that in the thermal balance the ions are much hotter then the
electrons. That also implies a high disc thickness.
In the highly turbulent plasma the energy exchange
between ions and electrons may be strongly enhanced due to presence
of fluctuating electrical fields, where electrons and ions gain the
same energy. In such conditions difference of temperatures between ions and
electrons may be negligible. The theory of relaxation
in the turbulent plasma is not completed, but there are indications
to a large enhancement of the relaxation in presence of plasma turbulence,
in comparison with the binary collisions (Galeev and Sagdeev, 1983;
Quataert, 1997). \\[2mm]

{\it 2.3. Accretion disc structure from equations describing continuously
optically thin and thick regions}\\[1mm]

Equations of the disc structure
smoothly describing transition between
optically thick and optically thin disc, had been obtained
using Eddington approximation.
The expressions had been obtained (Artemova et al., 1996)
for the vertical energy
flux from the disk $F_0$, and radiation pressure in the symmetry plane

\begin{equation}
\label{ref11.15}
 F_{0}={2acT_0^4 \over 3\tau_{0}\Phi},\,\,\,
 P_{rad,0}={aT_0^4 \over 3\Phi}\left(1+{4 \over 3\tau_{0}}\right),
\end{equation}
where
  $\tau_{\alpha 0}=\kappa_p \rho h={1 \over 2}\kappa_p \Sigma_0$,
  $\tau_{*}=\left(\tau_{0}\tau_{\alpha 0}\right)^{1/2}$,
 $\Phi=1+{4 \over 3\tau_{0}}+{2\over 3\tau_{*}^2}$.
At $\tau_0 \gg \tau_* \gg 1$ we have (\ref{ref1.12}) from (\ref{ref11.15}).
In the optically thin limit $\tau_* \ll \tau_0 \ll 1$ we get

\begin{equation}
\label{ref11.17}
 F_{0}=acT_0^4 \tau_{\alpha 0}, \quad
 P_{rad,0}={2 \over 3}acT_0^4 \tau_{\alpha 0}.
\end{equation}
Using $F_0$ instead of $Q_-$ and equation of state
$P=\rho {\cal R} T+P_{rad,0}$,
the equations of accretion disc structure together with equation
$Q_+=F_0$,
with $Q_+$ from (\ref{ref1.11}),
have been solved numerically by Artemova et al. (1996).
Two solutions, optically thick and thin, exist
separately when luminosity is not very large. They intersect at
$\dot m=\dot m_b$ and there is no global solution for accretion disc at
$\dot m > \dot m_b$. It was concluded by Artemova et al. (1996)
that in order to obtain a global physically meaningful solution
at $\dot m > \dot m_b$, account of advectivion is needed. For the
calculated case $M_{BH}=10^8\;M_\odot$, $\alpha=1.0$ at $\dot m=\dot m_b$
luminosity of the accretion disk is less than the critical Eddington
one. \\[2mm]

{\bf 3. Accretion discs with advection}\\[1mm]

Standard model gives somewhat nonphysical behavior near the inner
edge of the accretion disc around a black hole, with a zero heat production
at the inner edge of the disk. It is
clear from physical ground, that in this case
the heat brought by radial motion of matter
along the accretion disc becomes important. In presence of
this advective heating (or cooling) term,
depending on the radial entropy $S$ gradient, written as
$Q_{adv}=\frac{\dot M}{2\pi r}T \frac{dS}{dr}$,
the equation of a heat balance is modified to

\begin{equation}
\label{ref3.2}
Q_+ + Q_{adv}=Q_-.
\end{equation}
In order to describe self-consistently the structure of the accretion disc
we should also modify the radial disc equilibrium, including pressure
and inertia terms

\begin{equation}
\label{ref3.3}
r(\Omega^2-\Omega_K^2)=\frac{1}{\rho}\frac{dP}{dr}-v_r\frac{dv_r}{dr}.
\end{equation}
Appearance of inertia term leads to transonic radial flow with a
singular point. Conditions of a continuous passing of the solution through
a critical point choose a unique value of the integration constant $j_{in}$.
First approximate solution for the advective disc structure have been
obtained by Paczy\'nski and Bisnovatyi-Kogan (1981).
Attempts to find a solution for advective
disc had been done by Abramovicz et al. (1988),
Matsumoto et al. (1984).
For moderate values of $\dot M$ a unique
continuous transonic solution was found, passing through singular points,
and corresponding to a unique value of $j_{in}$.
The number of critical point in the radial flow with the
gravitational potential $\phi_g$
(Paczy\'nski and Wiita, 1980)
$\phi_g = \frac{GM}{r-r_g}, \quad r_g=\frac{2GM}{c^2}$.
may exceed unity. Appearance of two critical
points for a radial flow in this potential was analyzed by Chakrabarti
and Molteni (1993).
Using of equations averaged over a thickness of the disc
changes a structure of hydrodynamic equations, leading to a position
of singular points not coinciding with a unit Mach number point.
\\[2mm]

{\bf 4. Amplification of the magnetic field at a spherical accretion}\\[1mm]

A matter flowing into a black hole
is usually magnetized. Due to more rapid increase
of a magnetic energy the
role of the magnetic field increases when matter flows inside.
It was shown by Schwartsman (1971),
that magnetic energy density $E_M$ approaches
a density of a kinetic energy $E_k$, and he proposed a hypothesis of
{\it equipartition} $E_M \approx E_k$,
supported by continuous annihilation of the magnetic field in a region of
the main energy production. This hypothesis is
usually accepted in the modern picture of accretion (Narayan and Yu, 1995).
Account of the heating of
matter by magnetic field annihilation was done by
Bisnovatyi-Kogan and Ruzmaikin (1974).
For a spherical accretion with ${\bf v}=(v_r,\, 0,\, 0)$ the equations
describing a field amplification in the
ideally conducting plasma reduce to
(Bisnovatyi-Kogan, Ruzmaikin, 1974)

\begin{equation}
\label{ref2.10}
\frac{d( r^2 B_r)}{dt}=0, \quad \frac{d(r v_r B_{\theta})}{dt}=0, \quad
\frac{d(r v_r B_{\phi})}{dt}=0,
\end{equation}
where
$\frac{d}{dt}= \frac{\partial}{\partial t}+v_r\frac{\partial}{\partial r}$
is a full Lagrangian derivative.
Consider a free fall case with $v_r=-\sqrt{\frac{2GM}{r}}$.
The initial condition problem is solved separately for
poloidal and toroidal fields. For initially uniform field
$B_{r0}=B_0\cos\theta,
\,\, B_{\theta 0}=-B_0\sin\theta$
we get the solution (Bisnovatyi-Kogan, Ruzmaikin, 1974)

\begin{equation}
\label{ref2.15}
B_r=\frac{B_0\cos\theta}{r^2}\Phi_1^{4/3},\,\,\,
B_{\theta}=-\frac{B_0\sin\theta}{\sqrt r}\Phi_1^{1/3},
\end{equation}
where
$\Phi_1=r^{3/2}+\frac{3}{2}t\sqrt{2GM}$.
The radial component of the field is growing most rapidly. It is $\sim r^{-2}$
for large times, $\sim t^{4/3}$ at given
small radius, and is growing with time everywhere.
For initially dipole magnetic field

$$B_{r0}=\frac{B_0\cos\theta}{r^3},
\,\, B_{\theta 0}=-\frac{B_0\sin\theta}{2 r^3}$$
\noindent
we obtain the following solution

\begin{equation}
\label{ref2.16}
B_r=\frac{B_0\cos\theta}{r^2}\Phi_1^{-2/3},\,\,\,
B_{\theta}=-\frac{B_0\sin\theta}{2\sqrt r}\Phi_1^{-5/3}.
\end{equation}
Here the magnetic field is decreasing everywhere with time, tending to
zero. That describes a pressing of a dipole magnetic field to a
stellar surface. The azimuthal stellar magnetic field if confined
inside the star. When outer layers of the star are compressing with
a free-fall speed, then for initial field distribution
$B_{\phi 0}=B_0 r^n f(\theta)$ the change of $B_{\phi}$ with time is
described by a relation
$\quad B_{\phi}=-\frac{B_0\,f(\theta)}{\sqrt r}\Phi_1^{n+1/3}$.\\[2mm]

{\bf 5. Two-temperature advective discs}\\[1mm]

In the optically thin accretion discs at low mass fluxes the density
of the matter is low and energy exchange between electrons and ions
due to binary collisions is slow. In this situation, due to different
mechanisms of heating and cooling for electrons and ions, they may
have different temperatures. First it was realized by Shapiro et al. (1976),
where advection was not included. It was
noticed by Narayan and Yu (1995),
that advection in this case is becoming
extremely important. It may carry the main energy flux into a black
hole, leaving rather low efficiency of the accretion up to $10^{-3}\,-\,
10^{-4}$ (advective dominated accretion flows - ADAF).
This conclusion is valid only when the effects, connected
with heating of matter by magnetic field annihilation are
neglected.

  In the ADAF solution the ion temperature is about a virial one
$kT_i \sim GMm_i/r$, what means that even at high initial angular
momentum the disc becomes thick, forming a quasi-spherical
accretion flow.
When energy losses by ions are low, some kind of a "thermo-viscous"
instability is developed, because heating increases a viscosity, and
viscosity increases a heating. Development of this instability
leads to formation of ADAF.

A full account of the processes, connected with a presence of magnetic
field in the flow, is changing considerably the picture of ADAF. It was
shown by Schwarzman (1971), that
in the region of the main energy production, the
condition of equipartition takes place, and
efficiency of a radiation
increase enormously from $\sim 10^{-8}$ up to $\sim 0.1$
due to magneto-bremstrahlung.
To support the condition of equipartition a
continuous magnetic field reconnection
and heating of matter due to Ohmic dissipation takes place. It was
shown by Bisnovatyi-Kogan and Ruzmaikin (1974), that due to Ohmic heating
the efficiency of a radial
accretion into a black hole may become as high as
$\sim 30\%$. The rate of the Ohmic heating in the condition of
equipartition was obtained in the form

\begin{equation}
\label{ref3.5}
T\frac{dS}{dr} = -\frac{3}{2}\frac{B^2}{8\pi \rho r}.
\end{equation}
In the supersonic flow of the radial accretion equipartition
was suggested in a form (Schwartsman, 1971)
$\frac{B^2}{8\pi} \approx \frac{\rho v_r^2}{2}=\frac{\rho GM}{r}$.
For the disc accretion equipartition
between magnetic and turbulent energy
was suggested by Shakura (1972), what reduces with account of
"alpha" prescription of viscosity to a relation
$\frac{B^2}{8\pi}  \sim \frac{\rho v_t^2}{2} \approx \alpha_m^2 P$,
where $\alpha_m$ characterizes a magnetic viscosity in a way similar to
the turbulent $\alpha$ viscosity. It was suggested by Bisnovatyi-Kogan and
Ruzmaikin (1976)
the similarity between viscous and magnetic Reynolds
numbers, or between turbulent and magnetic viscosity coefficients
$Re=\frac{\rho v l}{\eta}, \quad Re_m=\frac{\rho v l}{\eta_m}$,
where the turbulent magnetic viscosity $\eta_m$ is connected with a turbulent
conductivity $\sigma=\frac{\rho c^2}{4\pi \eta_m}$.
Taking $\eta_m=\frac{\alpha_m}{\alpha}\eta$, we get a turbulent conductivity

\begin{equation}
\label{ref3.10}
\sigma=\frac{ c^2}{4\pi \alpha_m h v_s}, \quad v_s^2=\frac{P_g}{\rho}
\end{equation}
in the optically thin discs. For the radial accretion the turbulent
conductivity may contain mean free path of a turbulent element $l_t$
in (\ref{ref3.10}) instead of $h$.
In ADAF solutions, where ionic temperature is of the order of the virial one
two above suggestions for magnetic equipartition almost coincide at
$\alpha_m \sim 1$.

The heating of the matter due to an Ohmic dissipation
may be obtained from the Ohm's law in radial accretion

\begin{equation}
\label{ref3.11}
T\frac{dS}{dr} =-\frac{ \sigma {\cal E}^2}{\rho v_r}
\approx -\sigma \frac{v_E^2 B^2}{\rho v_r c^2}
=-\frac{B^2 v_E^2}{4 \pi \rho \alpha_m v_r v_s l_t},
\end{equation}
what coincides with (\ref{ref3.5})
when $ \alpha_m=\frac{4 r v_E^2}{3 v_r v_s l_t}$, or $l_t
=\frac{4 r v_E^2}{3 v_r v_s \alpha_m}$.
Here a local electrical field strength
in a highly conducting plasma is of the
order of ${\cal E} \sim \frac{v_E B}{c}$, $v_E \sim v_t
\sim \alpha v_s$  for a radial accretion.

Equations for a radial temperature dependence in the accretion disc,
separately for the ions and electrons are written as

\begin{equation}
\label{ref3.12a}
{dE_i\over dt}-{P_i\over\rho^2}{d\rho\over dt}=
{\cal H}_{\eta i}+{\cal H}_{Bi}-Q_{ie}~,
\end{equation}
\begin{equation}
\label{ref3.12b}
{dE_e\over dt}-{P_e\over\rho^2}{d\rho\over dt}=
{\cal H}_{\eta e}+{\cal H}_{Be}+Q_{ie}-{\cal C}_{ff}-{\cal C}_{cyc}~,
\end{equation}
A rate of a viscous heating of ions ${\cal H}_{\eta i}$ is obtained from
(\ref{ref1.11}) as

\begin{equation}
\label{ref3.13}
{\cal H}_{\eta i}=\frac{2\pi r}{\dot M}Q_+ \,=\,\frac{3}{2}
\alpha\frac{v_K v_s^2}{r}, \quad {\cal H}_{\eta e} \le
\sqrt{\frac{m_e}{m_i}} {\cal H}_{\eta i}.
\end{equation}
Combining (\ref{ref1.3}),(\ref{ref1.14a}),(\ref{ref1.9}),
we get
\begin{equation}
\label{ref3.14}
v_r=\alpha\frac{v_s^2}{v_K {\cal J}}, \quad h=\sqrt{2}\frac{v_s}{v_K} r,\quad
\rho=\frac{\dot M}{4 \pi \alpha \sqrt{2}}\frac{v_K^2 {\cal J}}{r^2v_s^3},
\end{equation}
where $v_K=r\Omega_K$, ${\cal J}=1-\frac{j_{in}}{j}$.
The rate of the energy exchange between ions and
electrons due to the binary collisions was obtained by Landau (1937).
Neglecting
pair formation for a low density accretion disc, we may write an exact
expression for a pressure
$P_g=P_i+P_e=n_ikT_e+n_ekT_p = n_e k(T_e+T_i)$,
and an approximate expression  for an energy, containing a smooth
interpolation between nonrelativistic and relativistic electrons.
The bremstrahlung ${\cal C}_{ff}$ and magneto-bremstrahlung
${\cal C}_{cyc}$ cooling of maxwellian semi-relativistic electrons, with
account of free-bound radiation in nonrelativistic limit, may be written as
interpolation of limiting cases (Bisnovatyi-Kogan and Ruzmaikin, 1976).

In the case of a disk accretion there are several characteristic velocities,
$v_K$, $v_r$, $v_s$, and $v_t=\alpha v_s$,
 all of which may be used for determining "equipartition"
magnetic energy, and one characteristic length $h$.
Consider three possible choices of $v_B^2$=$v_K^2$, $v_r^2$, and
$v_t^2$  for scaling $B^2=4\pi \rho v_B^2$.
The expression for an Ohmic heating in the turbulent accretion disc also
may be written in different ways, using different velocities $v_E$ in the
expression for an effective electrical field
${\cal E}=\frac{v_E B}{c}$.
A self-consistency of the model requires, that expressions
for a magnetic heating of the matter ${\cal H}_B$, obtained from the
condition of stationarity of the flow (\ref{ref3.5}), and from the Ohm's
law (\ref{ref3.11}), should be identical.
It happens at
$\frac{\alpha}{{\cal J}\alpha_m}{v_E^2 \over v_r^2}=\frac{3\sqrt{2}}{4}$.
That implies $v_E \sim v_r \sim \frac{\alpha v_s^2}{v_K \sqrt{\cal J}} \simeq
\frac {v_t v_s}{v_K \sqrt{\cal J}} <v_t$.
In the advective models
${\cal J}$ is substituted by a function which is
not zero at the inner edge of the disc. The heating due to magnetic
field reconnection ${\cal H}_B$ in the equations (\ref{ref3.12a}),
(\ref{ref3.12b}) may be written
with account of (\ref{ref3.13}) as
${\cal H}_B=\frac{3}{16 \pi}\frac{B^2}{r \rho} v_r
    = \frac{1}{2{\cal J}}{\cal H}_{\eta i}\left(\frac{v_B}{v_K}\right)^2$.
At $v_B=v_K$ the expressions for viscous and magnetic heating are
almost identical. Observations
of the magnetic field reconnection in the solar flares show
(Tsuneta, 1996), that electronic heating prevails over the ionc one.
Transformation of the magnetic energy into
a heat is connected with the change of the magnetic flux, generation
of the vortex electrical field, accelerating the particles.

The equations (\ref{ref3.12a}), (\ref{ref3.12b})
have been solved by Bisnovatyi-Kogan and Lovelace (1997)
for nonrelativistic electrons, at $v_B$=$v_K$.
The combined heating of the electrons and ions were taken as
${\cal H}_e=(2-g){\cal H}_{\eta i},\quad {\cal H}_e =g{\cal H}_{\eta i}$.
The results of calculations for
$g=0.5 \div 1$ show that almost all energy of the electrons is radiated,
so the relative efficiency of the two-temperature, optically thin disc
accretion cannot become lower then 0.25.
Increase of the term $Q_{ie}$ due to plasma turbulence
may restore the relative efficiency to a value,
corresponding to the optically thick disc. \\[2mm]

{\bf 6. Discussion}\\[1mm]

Observational evidences for existence of black holes inside our Galaxy
and in the active galactic nuclei (Cherepashchuk, 1996; Ho, 1999)
make necessary to revise theoretical models of the disc accretion.
The improvements of a model are connected with account of
advective terms and more accurate treatment of the magnetic field effects.
Account of the effects connected with magnetic field annihilation
does not permit to make a relative efficiency of the accretion lower
then $\sim 0.25$ from the standard value.
Strong relaxation connected with the plasma turbulence
may increase the efficiency, making it close to unity.
 For explanation of underluminous galactic nuclei two possible
 ways may be suggested. One is based on a more accurate estimations
 of the accretion mass flow into the black hole, which could be
 overestimated. The second is based
 on existence of another mechanisms of the energy losses in the form of
 accelerated particles, like in the radio-pulsars, where these losses
 exceed strongly a radiation. This is very probable to happen in
a presence of a large scale magnetic field which may be also responsible for
a formation of the observed jets (Bisnovatyi-Kogan, 1999;
Blandford and Begelman, 1999).
We may  suggest, that
underlumilnous AGN loose main part of their energy to the formation of jets,
like in SS 433.
The search of the correlation between existence of jets and lack of
the luminosity could be very informative. \\[2mm]

{\it Acknowledgements.}
The author is grateful for partial support to RFBR, grant 99-02-18180,
and GPNT "Astronomy", grant 1.2.6.5.
\\[3mm]
\indent
{\bf References\\[2mm]}
Abramovicz M.A., Czerny B., Lasota J.P., \h
Szuszkiewicz E.: 1988, {\it ApJ}, {\bf 332}, 646.\\
Artemova I.V., Bisnovatyi-Kogan G.S., Bj\"ornsson G.,\h
Novikov I.D.: 1996, {\it ApJ}, {\bf 456}, 119. \\
Bisnovatyi-Kogan G.S.: 1999, in {\it "Observational \h
evidences for black holes in the universe"},\h
ed. S.Chakrabarti, Kluwer, p.1.\\
Bisnovatyi-Kogan G.S., Blinnikov S.I.: 1976, \h
{\it Pisma Astron. Zh.}, {\bf 2}, 489.\\
Bisnovatyi-Kogan G.S., Lovelace R.V.L.: 1997, {\it ApJL}, \h
{\bf 486}, L43. \\
Bisnovatyi-Kogan G.S., Lovelace R.V.L.: 1999, \h
{\it ApJ} (accepted), astro-ph/9902344.\\
Bisnovatyi-Kogan G.S., Ruzmaikin A.A.: 1974, \h
{\it Ap. and Space Sci.}, {\bf 28}, 45.\\
Bisnovatyi-Kogan, G.S., Ruzmaikin, A.A., 1976, \h
{\it Ap. and Space Sci.}, {\bf 42}, 401.\\
Blandford R.D., Begelman M.C.: 1999, \h
{\it Month. Not. R.A.S.}, {\bf 303}, 1.\\
Cannizzo J.K. et al.: 1998, {\it AAS Abstracts}, {\bf 192}, 4103C.\\
Chakrabarti S.K., Molteni D.: 1993, {\it ApJ}, {\bf 417}, 671.\\
Cherepashchuk A.M.: 1996, {\it Uspekhi Fiz. Nauk.}, \h
{\bf 166}, 809.\\
Di Matteo T., Fabian A.C., Rees, M.J. et al.: 1999,\h
{\it Month. Not. R.A.S.}, {\bf 305}, 492.\\
Galeev A.A., Sagdeev R.Z.: 1983, Chap. 6, in \h
{\it Handbook of Plasma Physics}, Vol. 1, eds. \h
Rosenbluth M.N., Sagdeev R.Z. (Amsterdam: \h
North-Holland Pub.), Ch. 6.1\\
Ho L.: 1999, in {\it "Observ. evidences for b. h. in the \h
Universe"}, ed. S.Chakrabarti, Kluwer, p.157.\\
Landau L.D.: 1937, {\it Zh. Exp. Theor. Phys}. {\bf 7}, 203.\\
Matsumoto R., Sato Sh., Fukue J., Okazaki A.T.: \h
1984, {\it Publ. Astron. Soc. Japan}, {\bf 36}, 71.\\
Narayan R., Yu I.: 1995, {\it ApJ}, {\bf 452}, 710.\\
Novikov I.D., Thorne K.S.: 1973, in {\it Black Holes} \h
eds. C.DeWitt, B.DeWitt (New York: Gordon \& \h
Breach), p.345\\
Paczy\'nski B., Bisnovatyi-Kogan G.S.: 1981, \h
{\it Acta Astron.}, {\bf 31}, 283.\\
Paczy\'nski B., Wiita P.J.: 1980, {\it A\&A}, {\bf 88}, 23.\\
Pringle J.E., Rees, M.J.: 1972, {\it A\&A}, {\bf 21}, 1.\\
Quataert E.: 1997, {\it astro-ph/9710127}.\\
Schwartsman V.F.: 1971, {\it Soviet Astron.}, {\bf 15}, 377.\\
Shakura N.I.: 1972, {\it Astron. Zh.}, {\bf 49}, 921; \h
(1973, Sov. Astron., 16, 756)\\
Shakura N.I.,  Sunyaev R.A.: 1973, {\it A\&A}, {\bf 24}, 337.\\
Shapiro S.L., Lightman A.P., Eardley D.M.: 1976, \h
{\it ApJ}, {\bf 204}, 187.\\
Tsuneta S.: 1996, {\it ApJ}, {\bf 456}, 840.\\
\end{document}